\documentclass[12pt]{article}
\usepackage{graphicx, epsfig}
\usepackage{amsmath}
\usepackage{latexsym}
\usepackage{amstext}
\usepackage{amssymb}
\usepackage{array}
\usepackage{multirow}
\usepackage{tikz}
\newcommand{\mb}[1]{ \mbox{\boldmath$#1$} }

\newcommand{\ds}{\displaystyle}
\newcommand{\beq}{\begin{eqnarray}}
\newcommand{\eeq}{\end{eqnarray}}
\newcommand{\beqq}{\begin{eqnarray*}}
\newcommand{\eeqq}{\end{eqnarray*}}
\newcommand{\p}{\partial}

\newcommand{\eps}{\varepsilon}
\newcommand{\x}{\mbox{\boldmath$x$}}

\newcommand{\y}{\mbox{\boldmath$y$}}

\newcommand{\n}{\mbox{\boldmath$n$}}

\newcommand{\w}{\mbox{\boldmath$w$}}

\font\bb=msbm10 at 12pt
\def\rR{\hbox{\bb R}}

\def\v{\mathbf{v}}

\begin{document}

\pagestyle{plain}
\begin{center}
{\large \textbf{{Reconstructing a point source from diffusion fluxes to narrow windows in three dimensions}}}\\[5mm]
U. Dobramysl$^{1}$, D. Holcman$^{2}$
\footnote{  $^1$ Cancer Research UK Gurdon Institute, University of Cambridge, United Kingdom $^2$  Group of data modeling and computational biology, IBENS-PSL Ecole Normale Superieure, Paris, France. }
\end{center}
\date{}
\begin{abstract}
We develop a computational approach to locate the source of a steady-state gradient of diffusing particles from the fluxes through narrow windows distributed either on the boundary of a three dimensional half-space or on a sphere. This approach is based on solving the mixed boundary stationary diffusion equation with the Neumann-Green's function and matched asymptotic.  We compute the probability fluxes and develop a highly efficient analytical-Brownian numerical scheme. This scheme accelerates the simulation time by avoiding the explicit computation of Brownian trajectories in the infinite domain. Our derived analytical formulas agree with the results obtained from the fast numerical simulation scheme. Using the analytical representation of the particle fluxes, we show how to reconstruct the location of the point source. Furthermore, we investigate the uncertainty in the source reconstruction due to additive fluctuations present in the fluxes. We also study the influence of various window configurations (cluster vs uniform distributions) on recovering the source position. Finally, we discuss possible applications in cell biology.
\end{abstract}
keywords: Narrow escape; diffusion; mixed-boundary value; Green's function; Brownian simulations; inverse problem

\section{Introduction}
\begin{figure}
    \centering
    \includegraphics[scale=0.4]{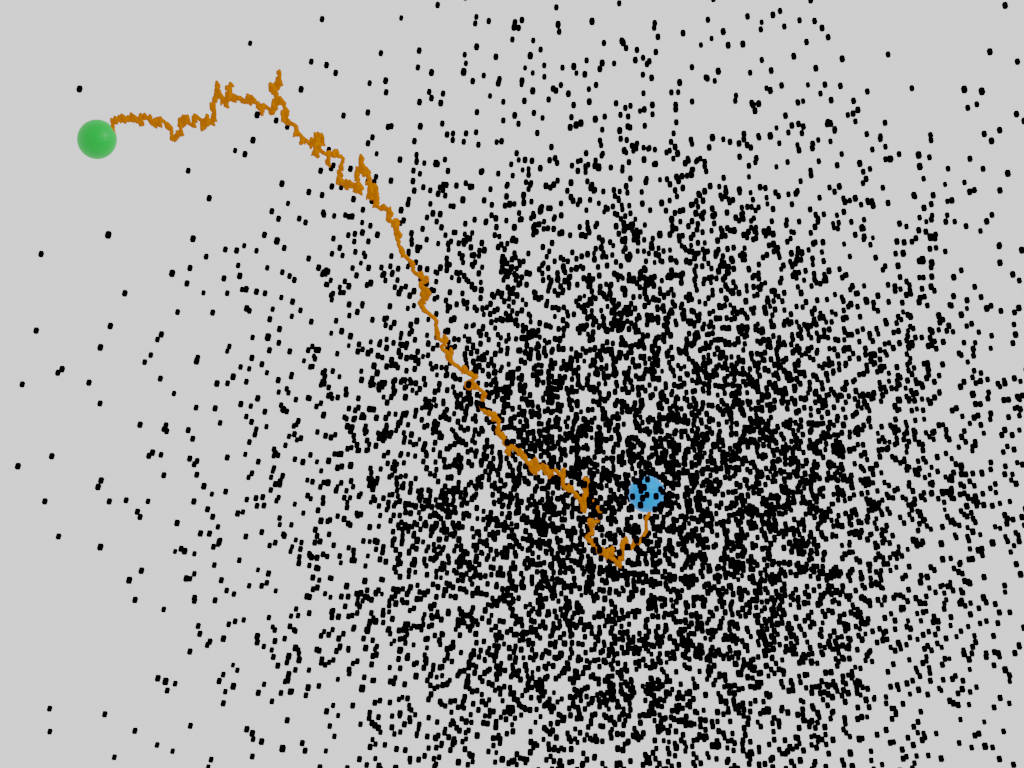}
    \caption{{\bf Scheme of a cell in a gradient}. A cell (green) is embedded in a gradient of cues (black) and has to find a target source (blue), using an optimal path, that depends on the gradient ralization.}
    \label{fig:figure0}
\end{figure}

We present a general computational approach to recover the position of a point source, which emits stochastic particles, from the steady-state fluxes collected at narrow windows. These windows are located on a planar surface or a ball. This approach is motivated by the following biological question: how a cell can determine the point source generating a molecular gradient in three dimensions. Indeed, in order to navigate a cell embedded in a tissue has to determine its position relative to guidance points. For example, bacteria are finding a local gradient source of diffusing molecules and are basing their movement decisions on this information, as illustrated in Fig. \ref{fig:figure0}.\\
In general, sensing a molecular gradient is a key process in cell biology and crucial for the detection of a concentration that can transform  positional information into specialization and differentiation \cite{wolpert1996one,Malherbe,Kasatkin2,kasatkin2008morphogenetic}. During neural development, the tip of the axonal projection - the growth cone - uses the concentration of morphogens \cite{reviewREingruber} to decide whether to continue moving or to stop, to turn right or left. Bacteria and spermatozoa are able to orient themselves in a chemotaxis gradient \cite{Heinrich3,Kaupp1}.\\
Models in the current literature usually consider the external gradient and are mostly focusing on computing the flux to a test absorbing ball using \emph{uniform} boundary conditions \cite{BergPurcell,Endres2008} which is sufficient to detect a gradient direction, but not the source position. The spatial distribution of receptors on the cell surface that bind cue molecules can also influence the fluxes of diffusing particles. Generally, these binding events occurs at fast timescales compared to diffusion or cell movement and receptors report their binding state - i.e. the diffusive flux to the receptor - to the interior of the cell via biochemical signalling cascades. \\
Based on the information about the diffusive receptor fluxes, we would like to ask the question whether or not the location of the gradient can be found? If yes, what is the minimum number of receptors needed to do so? Motivated by the principles of triangulation during navigation, not unlike how the Global Positioning System allows the positioning via the signal from at least three satellites, we use a diffusion model to compute the fluxes through narrow windows distributed on the boundary of half-space and a ball in three dimensions.\\
We recall that computing the fluxes of Brownian particles to small targets located on the surface of a domain is the goal of the Narrow Escape Theory \cite{SIREV-biol,coombs,ward1,ward2,ward3,JPA2017,Holcman2015}. We simplify here the cell geometry as a sphere containing small targets on its surface (receptors). A diffusing molecule can find one of these receptors, leading to receptor activation. We neglect the binding or interaction time, so that the windows are considered to be purely absorbing. The receptor activation can mediate further cellular transduction that signals the external environment to the interior of a cell. When a cell has to compare the difference between the fluxes from the left and the right, the asymmetry of fluxes at the receptors should be kept so that this difference creates a local signal to prevent the loss due to homogenization inside the cell. Indeed, if the signal which can be the concentration of second messengers or surface molecules is spread uniformly, then the direction of the gradient is lost. We study here the effect of receptor distribution on their fluxes and we do not replace them by a homogenized boundary condition that would prevent the ability to detect flux differences between receptors at different locations.\\
In this manuscript, we compute the flux of molecules to small targets located on the surface of cell. Asymptotic computations and numerical simulations reveal the influence of parameters such as the cell geometry, the distribution of target receptors and possible cooperativity on the recovery of the location of the source. We show that it is possible to recover the source of a gradient with already three receptors, while sensing of the mean concentration level can be achieved with two only. Many recent hybrid algorithms to enhance computational effiency in microscopic diffusion have been introduced in the last decade~\cite{Flegg2011,Franz2013,Smith2018}. The novel aspect in this manuscript is the asymptotic solution of the Laplace's equation in infinite domains based on match asymptotic in three dimension. The mean passage time to a small hole, becomes infinite in an unbounded domain due to the long excursion of Brownian trajectories to infinity. In addition, particles can escape to infinity before hitting the narrow windows. This difficulty also plagues simulations, which is resolved here by introducing a new scheme. Indeed, the fluxes are not directly from entire Brownian trajectories generated in the entire space, but only from fragments generated very close to the domain of interest, avoiding the inefficient computation of the flux from long trajectories. We show that the analytical formulas and numerical simulations are in very good agreements.  Finally,  to quantify the uncertainty associated to the source recovery, we introduce a novel coordinate systems,  defined by the ensemble of three fluxes. This coordinate system allows us to define the possible positions and the volume where the source is located. To conclude, we find that in dimension three, adding receptors lead to a faster than exponential increase of the precision of the source recovery.

\section{Model of diffusing particles to absorbing holes in three dimensions}
\subsection{Steady-state Laplace equation as a mixed-boundary value problem}
In this section, we present a generic model to compute the distribution of fluxes between several absorbing windows located on the surface of a domain. The model consists of a steady-state source located at position $x_0$ releasing independent Brownian particles that can move in free space, but cannot penetrate a bounded domain $\Omega$ (which will either be a ball of radius $R$ or the half-space in the negative $z$ direction). The boundary $\p\Omega$ contains $N$ small and disjoint absorbing windows, $\partial\Omega_{\varepsilon_j}\ (j=1,\ldots,N)$ each of area $|\partial\Omega_{\eps_j}|=O(\eps^2)$, where the radius $\eps$ is small. We assume that the windows are sufficiently far apart to avoid non-linear effects~\cite{Holcman2008_1}. The total absorbing boundary is given by
\beq
\p\Omega_a= \cup_1^{N} \partial\Omega_{\varepsilon_j}.
\eeq
The remaining boundary surface is reflective $\p\Omega_r=\p\Omega-\p\Omega_a$ for the diffusing particles. Note that other models are possible, for example instead of purely absorbing boundary conditions one can consider partially absorbing (Robin) boundary conditions.

To compute the fluxes, we use the transition probability density $p(\x,t\,|\,\x_0)$ to find a particle at position $\x$ at time $t$, when it started at position $\x_0$. It is the solution of
\begin{align}\label{IBVP}
\frac{\p p(\x,t\,|\,\x_0)}{\p t} =&D \Delta p(\x,t\,|\,\x_0)\quad\hbox{\rm for } \x,\x_0 \in \rR^3-\Omega,\\
p(\x,0\,|\,\y)=&\delta(\x-\x_0)\quad \hbox{\rm for } \x,\x_0 \in \rR^3-\Omega\nonumber\\
\frac{\p p(\x,t\,|\,\x_0)}{\p \n} =&0\quad \hbox{\rm for } \x \in\p\Omega_r, \y\in\Omega\nonumber\\
p(\x,t\,|\,\x_0)=&0\quad \hbox{\rm for } \x \in \p\Omega_a,  \x_0\in \rR^3-\Omega,\nonumber
\end{align}
where D is the diffusion coefficient. The steady-state gradient $P_0$ is obtained by resetting a particle after it disappears through a window \cite{Schuss:Book}. It is given as the solution of the mixed-boundary value problem
\beq
D\Delta P_0(\x) & = & -\delta_{P_0} \hbox{ for } \x\, \in \,\rR^3 -\Omega \label{eqDP3b}\\
\frac{\p P_0}{\p \n}(\x) & = & 0\hbox{ for } \x \,   \in\, \p\Omega_r \\
P_0(\x) & = & 0 \hbox{ for } \x \,   \in \p\Omega_a.
\eeq
Our goal is to compute the probability fluxes associated to $P_0$ on each individual windows $\Omega_{\eps_j}$.  As we shall see, the fluxes depend on the specific window arrangement and the domain $\Omega$. Note that when $Q$ particles are injected per unit of time, the steady-state fluxes are computed from
\beq
-D\Delta P_0(\x) & = & Q\delta(\x-\x_0) \;\;\text{for}\;\; \x\, \in \,\rR^3 \label{eqDP1}
\eeq
The parameter $Q>0$ can be calibrated to a fixed number of particles located in a volume. At infinity, the density $P_0(\x)$ has to tend to zero in three dimensions. More complex domains could be studied if their associated Green's function can be found.

\subsection{Computing the fluxes of Brownian particles to small windows in half--space} \label{s:HP}
To compute the fluxes to narrow windows located on the plane $\rR_2$, when the Brownian particles can evolve in $\rR_{+}^3$, we will use the method of matched asymptotics. In the following, we set the diffusion coefficient to one, $D=1$.  We start by constructing a general solution of equation~\ref{eqDP3b} using the Green's function:
\beq
\Delta G_0(\x) & = & -\delta_{x_0} \hbox{ for } \x\, \in \,\rR^3  \label{eqDP3plus}\\
\frac{\p G_0}{\p \n}(\x) & = & 0\hbox{ for } \x \,   \in\, \rR_{+}^3.
\eeq
In three dimensions, the solution that tends to zero at infinity is
\beq
G(\x,\x_0) = \frac{1}{4\pi}\left(\frac1{|\x-\x_0|}  + \frac1{|\x - \overline{\x}_0|}\right).\label{eqDP33}
\eeq
where $\overline{\x}_0$ is the mirror image of $\x_0$ with respect to the plane at $z=0$. The function $w= P_0-G_0$ is the solution of
\beq
\Delta w & = & 0 \hbox{ for } \x\, \in \,\rR^3 \\
\frac{\p w}{\p \n}(\x) & = & 0\hbox{ for } \x \,   \in\, \rR_{+}^3 \\
w(\x) & = & \alpha_i \hbox{ for } \x \,   \in \Omega_{\eps_i}, \, i=1..N,
\eeq
where we consider that the windows $\Omega_{\eps_i}$ are circular and centered around the point $x_i$. We also assume that $\eps_i$ is small enough such that we can approximate the Green's function as being constant over the window:
\beq
\alpha_i= -G_0(x_i,x_0).
\eeq
We construct the solution of $w$ using the elementary solution $w_c$ of the boundary layer equation near each window $\Omega_{\eps_i}$
\beq
{\cal L} w_c &\equiv&\, {w_c}_{\eta\eta} + {w_c}_{s_1s_1} + {w_c}_{s_2s_2}
=0\hspace{0.5em}\mbox{for}\ \eta\geq 0, \ -\infty<s_1,s_2<\infty\label{5:wc_1}\\
\partial_\eta w_c &=& \,0\hspace{0.5em}\mbox{for}\
\eta=0,\ s_1^2+s_2^2\geq \eps_j^2, \quad   w_c = 1 \hspace{0.5em}
\mbox{for} \ \eta=0, \ s_1^2+s_2^2 \leq \eps_j^2\label{5:wc_2}\\
w_c&\to&\,0\hspace{0.5em}\mbox{for}\ \rho=\eps^{-1}|\x-\x_j|\to\infty.
\label{mcdn}
\eeq
The boundary value problem (\ref{5:wc_1}), (\ref{5:wc_2}) with the matching
condition (\ref{mcdn}) is the well-known electrified disk problem in
electrostatics (cf.~\cite{Jackson}), which has the solution
 \beq
 w_{c} = \frac{2}{\pi} \int\limits_{0}^{\infty} \frac{\sin\mu}{\mu} \, e^{-\mu
\eta/\eps_j} \, J_{0}\left( \frac{\mu \sigma}{\eps_j} \right) \, d\mu = \frac{2}{\pi}
\sin^{-1}\left(\frac{\eps_j}{L} \right), \label{5:wcsol_1}
 \eeq
where $\sigma \equiv (s_1^2 + s_2^2)^{1/2}$. The symbol $J_{0}(z)$ is the
Bessel function of the first kind of order zero, and $L=L(\eta,\sigma)$ is
defined by
 \beq
 L(\eta,\sigma) \equiv  \frac{1}{2} \left(  \left[ (\sigma + \eps_j)^2 + \eta^2
\right]^{1/2}  +   \left[ (\sigma - \eps_j)^2 + \eta^2 \right]^{1/2} \right) .
  \label{5:wcsol_2}
\eeq
The far-field behavior of $w_c$ in (\ref{5:wcsol_1}) is given by
\beq
w_{c} \sim \frac{2\eps_j}{\pi} \left[ \frac{1}{\rho} + \frac{\eps_j^2}{6}
 \left( \frac{1}{\rho^3} - \frac{3\eta^2}{\rho^5} \right) + \cdots
  \right]\hspace{0.5em}\mbox{as} \quad \rho \to \infty , \label{5:wcff}
\eeq
which is uniformly valid in $\eta$, $s_1$, and $s_2$.  {Thus (\ref{5:wcff}) gives the far-field expansion of $w_0$ as}
\beq
w_0 \sim v_0 \left( 1 - \frac{c_j}{\rho} + O(\rho^{-3}) \right)
\hspace{0.5em} \mbox{for} \ \rho \to \infty, \ c_j=\frac{2\eps_j}{\pi},
\label{5:w0ff}
\eeq
where $c_j$ is the electrostatic capacitance of the circular disk of radius
$\eps_j$.\\
In the half-plane situation, we can write the general solution using the solution $w_{c_i}$ for window $\Omega_{\eps,i} $ as the linear combination
\beq \label{solution}
w(x) =\sum_{i=1}^{n} a_i \alpha_i w_{c_i}(x).
\eeq
where the coefficients $a_i$ have to be determined. There are found using the absorbing boundary conditions on each window:
\beq
\alpha_j=\sum_{i=1}^{n} a_i \alpha_i w_{c_i}(x_j), \hbox{ for } j=1..n.
\eeq
By definition, $w_{c_i}(x_i)=1$.  Thus using the matrix
\beq\mb{M}= \left( \begin{array}{cccc}
\alpha_1 & \alpha_2 w_{c_2}(x_1) &... & \alpha_n w_{c_n}(x_1)  \\
&&&\\
\alpha_1 w_{c_1}(x_2) & .&.&.\\
&&&\\
.&. &.&.\\
&&&\\ \alpha_1 w_{c_1}(x_n)& ...&.& \alpha_n
\end{array}\right)\label{M}
 \eeq
and the approximation that for windows sufficiently far apart with the same radius $\eps$,
\beq
w_{c_i}(x_j)\approx \frac{2 \eps \alpha_i}{\pi |x_0-x_j|}
\eeq
we can now derive a Matrix equation. To this end, we decompose $\mb{M}$ as
\beqq
\mb{M}=\mb{\Delta}_{\alpha}+ \frac{2\eps}{\pi} \mb{A},
 \eeqq
where $\mb{\Delta}_{\alpha}$ is the diagonal matrix
\beq
\mb{\Delta}_{\alpha}= \left( \begin{array}{cccc}
\alpha_1 & 0 &... & 0  \\
&&&\\
0 & .&.&.\\
&&&\\
.&. &.&.\\
&&&\\ 0& ...&.& \alpha_n
\end{array}\right)\label{M2}
 \eeq
and $\mb{A}$ contains the off-diagonal terms:
\beq\mb{M}= \left( \begin{array}{cccc}
0 & \alpha_2 w_{c_2}(x_1) &... & \alpha_n w_{c_n}(x_1)  \\
&&&\\
\alpha_1 w_{c_1}(x_2) & .&.&.\\
&&&\\
.&. &.&.\\
&&&\\ \alpha_1 w_{c_1}(x_n)& ...&.& 0
\end{array}\right).\label{M3}
\eeq
Writing $\tilde{\mb{\alpha}}$ and $\tilde{\mb{a}}$ for the vectors containing the $\alpha_j$ and the $a_j$ respectively, equation \eqref{solution} becomes
\beq \label{sysMatrix}
\left(\mb{\Delta}_{\alpha}+ \frac{2\eps}{\pi} \mb{A}\right) \tilde{\mb{a}}=\tilde{\mb{\alpha}}.
 \eeq
This can be inverted as the following convergent series
\beq \label{formal1s}
\tilde{\mb{a}}= \left( \mb{1}_M+ \frac{2\eps}{\pi} \mb{\Delta}_{\alpha}^{-1} \mb{A}\right)^{-1} \mb{\Delta}_{\alpha}^{-1} (\tilde{\mb{\alpha}}) =-\sum_{k=0}^{\infty} \ds(- \frac{2\eps}{\pi} \mb{\Delta}_{\alpha}^{-1}\mb{A})^k
\mb{\Delta}_{\alpha}^{-1} (\tilde{\mb{\alpha}}).
 \eeq
Relation \ref{formal1s} is the formal solution for the coefficients $a_i$ in the asymptotic solution \ref{solution}. Finally, we recall that the flux through each window is given by
\beq\label{totalflux}
\Phi_i=\int_{\Omega_{i}}\frac{\p w}{\p \n}(\y) dS_{\y} =4 \eps \pi a_i \alpha_i.
\eeq
\subsection{Explicit expression in the cases of $n=1,2$ and $3$ windows in the $z=0$ plane} \label{ss:explicit}
We now compute the fluxes for one, two and three windows. When there is only one window, the asymptotic representation of solution \ref{solution} is
\beq
P_0(x)=G_0(x)-G_0(x_1)\frac{2}{\pi}
\sin^{-1}\left(\frac{a_j}{L(x)} \right),
\eeq
where $L(x)$ is defined by \ref{5:wcsol_2}. Therefore, we arrive at
\beq
\int_{\Omega_{i}}\frac{\p P_0}{\p \n}(\y) dS_{\y} =\int_{\Omega_{i}}\frac{\p G_0}{\p \n}(\y) dS_{\y} -G_0(x_1) \int_{\Omega_{i}}\frac{\p w_c(\y)}{\p \n}(\y) dS_{\y}.
\eeq
By definition $\int_{\Omega_{\eps}}\frac{\p G_0}{\p \n}(\y) dS_{\y} =0$ and thus the probability flux is given by
\beq
\Phi_{\eps}=\int_{\Omega_{\eps}}\frac{\p P_0}{\p \n}(\y) dS_{\y} =\frac{2}{\pi}\frac{\eps}{|x_0-x_1|}.
\eeq
We conclude that when the flux $\Phi_{\eps}$ is given, the ensemble of possible positions $\x_0$ is a sphere centered around $x_1$ and of radius $\frac{2}{\pi}\frac{\eps}{\Phi_{\eps}}$.

We now consider the case of two windows centered at $x_1$ and $x_2$, for which the solution \ref{solution} is now
\beq \label{solution2}
w(x) =a_1 \alpha_1 w_{x_1}(x)+a_2 \alpha_2 w_{x_2}(x).
\eeq
In this case, the solution of system \ref{sysMatrix} is an elementary 2 by 2 matrix and is given by
\beq
a_1&=&\frac{1-d_{\eps}\alpha_b/\alpha_a}{1-d_{\eps}^2}\\
a_2&=&\frac{1-d_{\eps}\alpha_a/\alpha_b}{1-d_{\eps}^2},
\eeq
where $d_{\eps}=\frac{2 \eps}{\pi |x_1-x_2|}$. Thus from relation \ref{totalflux}, the flux at each window can be computed explicitly:
\beq
\label{eq:halfspacefluxn=2}
\Phi_1=\int_{x_1+\Omega_{\eps}}\frac{\p P}{\p \n}(\y) dS_{\y} =4 \eps \pi a_1 \alpha_a.\\
\Phi_2=\int_{x_2+\Omega_{\eps}}\frac{\p P}{\p \n}(\y) dS_{\y} =4 \eps \pi a_2 \alpha_b.
\eeq
Hence we obtain the explicit representation:
\beq \label{twowindows}
\Phi_1=\frac{2\eps}{\pi|x_1-x_0|} \ds \left( \frac{1- \ds\frac{2 \eps |x_2-x_0| }{\pi |x_1-x_2||x_1-x_0|}}{1-\ds(\frac{2 \eps}{\pi |x_1-x_2|})^2}\right)\\
\Phi_2=\frac{2\eps}{\pi|x_2-x_0|} \ds \left( \frac{1- \ds \frac{2 \eps |x_1-x_0| }{\pi |x_1-x_2||x_2-x_0|}}{1-\ds (\frac{2 \eps}{\pi |x_1-x_2|})^2}\right).
\eeq
To conclude, when the two fluxes $\Phi_1$ and $\Phi_2$ are given, the possible position for the source from equation \ref{twowindows} is located at the intersection of two spheres and thus we are left with a one dimensional curve.

We now consider three windows located at $x_1$, $x_2$ and $x_3$. The solution \ref{solution} is now
\beq \label{solution2_1}
w(x) =a_1 \alpha_1 w_{x_1}(x)+a_2 \alpha_2 w_{x_2}(x)+a_3 \alpha_3 w_{x_3}(x)
\eeq
Inverting the matrix $\ref{sysMatrix}$, we obtain the explicit representation
\beq\label{exactN=3}
a_1&=&\frac{1-d_{23}^2+\frac{\alpha_2}{\alpha_1}(d_{13}d_{23}-d_{12})+\frac{\alpha_3}{\alpha_1}(d_{12}d_{23}-d_{13})}{1-\Delta^2}\\
a_2&=&\frac{1-d_{13}^2+\frac{\alpha_1}{\alpha_2}(d_{13}d_{23}-d_{12})+\frac{\alpha_3}{\alpha_2}(d_{12}d_{13}-d_{23})}{1-\Delta^2}\\
a_3&=&\frac{1-d_{12}^2+\frac{\alpha_1}{\alpha_3}(d_{12}d_{23}-d_{13})+\frac{\alpha_2}{\alpha_1}(d_{12}d_{13}-d_{23})}{1-\Delta^2},
\eeq
where $d_{ij}=\frac{2 \eps}{\pi |x_i-x_j|}$ and $\Delta^2=d_{12}^2+d_{13}^2+d_{23}^2+2d_{12}d_{13}d_{23}$. Plugging \ref{exactN=3} into \ref{totalflux} and expanding to second order in $\eps$ yields the expansion of the fluxes with respect $\eps$
\beq
\label{fluxesN=3}
\Phi_1&=&\frac{2\eps}{\pi}\frac{1}{|x_1-x_0|}-4\eps^2\left[\frac{1}{|x_2-x_0|}\frac{1}{|x_1-x_2|}+\frac{1}{|x_3-x_0|}\frac{1}{|x_1-x_3|}\right]+O(\eps^3)\\
\Phi_2&=&\frac{2\eps}{\pi}\frac{1}{|x_2-x_0|}-4\eps^2\left[\frac{1}{|x_1-x_0|}\frac{1}{|x_1-x_2|}+\frac{1}{|x_3-x_0|}\frac{1}{|x_2-x_3|}\right]+O(\eps^3)\\
\Phi_3&=&\frac{2\eps}{\pi}\frac{1}{|x_3-x_0|}-4\eps^2\left[\frac{1}{|x_1-x_0|}\frac{1}{|x_1-x_3|}+\frac{1}{|x_2-x_0|}\frac{1}{|x_2-x_3|}\right]+O(\eps^3).
\eeq
To conclude, it is now clear that with three windows with the values of the fluxes $(\Phi_1,\Phi_2,\Phi_3)$ given, we have three equations for the three unknown coordinates of the source position. The solution is uniquely given as the intersection of three spheres (to order $\eps$). This solution defines $x_0$ with the flux-coordinates $x_0(\Phi_1,\Phi_2,\Phi_3)$.
\subsection{Computing the fluxes of Brownian particles to small targets on the surface of a ball} \label{s:ball}
In this section, we compute the flux to narrow windows located a three dimensional ball $B_a$ of radius $a$, when Brownian particle are release at a position $\x_0$ outside the ball. The fluxes are computed from the solution of the associated Laplace's equation
\beq
D\Delta P_0(\x) & = & -\delta_{\x_0} \hbox{ for } \x\, \in \,\rR^3 -B_a \label{eqDP3bb}\\
\frac{\p P_0}{\p \n}(\x) & = & 0\hbox{ for } \x \,   \in\, \p B_a-S(\eps) \\
P_0(\x) & = & 0 \hbox{ for } \x \,   \in \Sigma_a=S_1(\eps) \cup..\cup S_n(\eps)
\eeq
where $S_k(\epsilon)$ are non-overlapping windows of radius $\eps$ located on the surface of the ball and centered around the point $\x_k$. We have the additional condition at infinity:
\beq
\lim_{|\x| \rightarrow \infty}  P_0(\x) =0.
\eeq
Following the first step described in subsection \ref{s:HP}, we compute the difference $w= P_0-\tilde N$, where $\tilde N$ is the Neumann-Green function for the external Ball defined in \ref{Neumann}. It is the solution of
\beq \label{eqfdt}
\Delta w & = & 0 \hbox{ for } \x\, \in \,\rR^3 -B_a \\
\frac{\p w}{\p \n}(\x) & = & 0\hbox{ for } \x \,   \in\, \Sigma_a\\
w(\x) & = & \alpha_i \hbox{ for } \x \,   \in S_i(\eps), \, i=1..N
\eeq
where we again consider that the windows $\Omega_{\eps_i}$ to be small enough such that we can approximate the Green's function as a constant
\beq
\alpha_i= -N(\x_i,\x_0).
\eeq
To solve \ref{eqfdt}, we use Green's identity over the large domain $\Omega$,
\beq
\int_{\Omega}\left( \mathcal{N}(\x,\x_0)  \Delta w(\mathbf{x})-\tilde{p}(\mathbf{x}) \Delta \mathcal{N}(\x,\x_0) \right)d\mathbf{x}= \int_{\p \Omega} \left( \mathcal{N}(\x,\x_0) \frac{\p w(\mathbf{x})}{\p n} -w(\x) \frac{\p \mathcal{N}(\x,\x_0)}{\p n}\right).
\eeq
Using expressions \ref{eqfdt} and \ref{Neumann}, we obtain
\beq
w(\x)=\sum_{k} \int_{S_k(\eps)} \mathcal{N}(\x,\x_k) \frac{\p w(\x)}{\p n} dS_{\x},
\eeq
where we use that the unbounded part of the surface integral in $\Omega$ converges to zero at infinity due to the decay condition \ref{decay}. We recall that the flux to an absorbing hole \cite{Holcman2015} is
\beq
\frac{\p P_0}{\p \n}(\y) =\frac{A_i}{\sqrt{\eps^2-r^2}}, \hbox{ for } \y \in S_k(\eps).
\eeq
To compute the unknown constants $A_i$, we use the Dirichlet condition at each window
\beq \label{matrixball}
\alpha_q&=&\int_{S_q(\eps)} \mathcal{N}(\x_q,\x) \frac{\p w(\x)}{\p n} dS_{\x}+\sum_{k\neq q} \int_{S_k(\eps)} \mathcal{N}(\x_q,\x_0) \frac{\p w(\x)}{\p n} dS_{\x},\\
&=&\mathcal{N}(\x_q,\x_k) \int_{S_q(\eps)} \frac{\p w(\x)}{\p n} dS_{\x}+\sum_{k\neq q} \int_{S_k(\eps)} \mathcal{N}(\x_q,\x) \frac{\p w(\x)}{\p n} dS_{\x}.
\eeq
Using Neumann's representation \ref{Neumann2} for the singularity located on the surface of the disk, the first integral term in expression \ref{matrixball} \cite{Lagache2017} yields:
\beqq
\int_{S_q(\eps)} \mathcal{N}(\x_q,\x) \frac{\p w(\x)}{\p n} dS_{\x}
&\approx &\int_{0}^{\epsilon} \left(\frac{g_0^i}{\sqrt{\epsilon^2-s^2}}+f_i(s)\right)\left(\frac{1}{2\pi Ds}+ \frac{1}{4\pi a }\log\left(\frac{s}{2a+s}\right)+O(1)\right) 2\pi s ds \nonumber \\
&=& A_k\left(\frac{\pi}{2}+\frac{\epsilon}{2a} \log\left(\frac{\epsilon}{a}\right)+B_k\epsilon \right),\nonumber
\eeqq
where $B_k$ is a constant term appearing in the third order expansion of the Green's function \cite{Lagache2017}. For the second term, we recall that
\beq
\int_{0}^{\epsilon}\frac{A_k}{\sqrt{\epsilon^2-s^2}} 2\pi s ds=2\pi \eps A_k,
\eeq
and for $k=1..N$ obtain the relations
\beq \label{matrixball2}
\alpha_q &=& 2\pi \eps \sum_{k\neq q} A_k \mathcal{N}(\x_q,\x_k)+A_q\left(\frac{\pi}{2}+\frac{\epsilon}{2a} \log\left(\frac{\epsilon}{a}\right)+B_k\epsilon \right),
\eeq
which can be written in Matrix form:
\beq \label{sysMatrix2}
[\mb{\tilde M}]{\mb{ \tilde A}}=\tilde{\mb{\alpha}}.
\eeq
We decompose $[\mb{\tilde M}]$ as
\beqq
[\mb{\tilde M}]=\mb{\Delta}+ \frac{2\eps}{\pi} \mb{N},
 \eeqq
where
\beq\mb{N}= \left( \begin{array}{cccc}
0 & \mathcal{N}(\x_1,\x_2) &... & \mathcal{N}(\x_1,\x_n)  \\
&&&\\
\mathcal{N}(\x_1,\x_2) & .&.&.\\
&&&\\
.&. &.&.\\
&&&\\ \mathcal{N}(\x_1,\x_n)& ...&.& 0
\end{array}\right),\label{NMat}
 \eeq
and
\beqq
\mb{\Delta}=\theta_{\eps}\mb{I}.
\eeqq
Here,
$\theta_{\eps}=\left(\frac{\pi}{2}+\frac{\epsilon}{2a} \log\left(\frac{\epsilon}{a}\right)+B\epsilon \right)$
and
\beq
\tilde{\mb{\alpha}}=
\left( \begin{array}{cc}
\alpha_1 & \\
.&\\
.&\\
\alpha_n
\end{array}\right). \label{alpha2}\,
{\mb{ \tilde A}}= \left( \begin{array}{cc}
A_1 & \\
.&\\
.&\\
A_n
\end{array}\right).
\label{alpha2_1}
\eeq
By inverting the matrix, we obtain the solution for the flux constants
\beq \label{formals}
\tilde{\mb{A}}= \left( \theta_{\eps}I+ \frac{2\eps}{\pi} \mb{\Delta}_{\alpha}^{-1} \mb{A}\right)^{-1} \mb{\Delta}_{\alpha}^{-1} (\tilde{\mb{\alpha}}) =-\sum_{k=0}^{\infty} \ds(- \frac{2\eps}{\pi} \mb{\Delta}_{\alpha}^{-1}\mb{A})^k \mb{\Delta}_{\alpha}^{-1} (\tilde{\mb{\alpha}}).
\eeq
Finally, the flux to each window is, to first approximation,
\beq \label{flux3dd}
\Phi_k=\int_{S_q(\eps)} \frac{\p P(\x)}{\p n} dS_{\x}=\int_{S_q(\eps)} \frac{\p w(\x)}{\p n} dS_{\x}=2\pi A_k=\theta_{\eps}^{-1} (\alpha_k -\frac{2 \pi \eps}{\theta_{\eps}}\sum_{j\neq k}\mathcal{N}(\x_q,\x_k)\alpha_k )+O((\frac{2 \pi \eps}{\theta_{\eps}})^2).
\eeq
System \ref{flux3dd} can be solved numerically to recover the flux solution ($A_k$ and $\Phi_k$) depending on the source position $\x_0$ and the distribution of the windows $\x_1,..\x_n$.
\section{Hybrid stochastic simulations}
\begin{figure}[http!]
    \centering
    \includegraphics[scale=0.8]{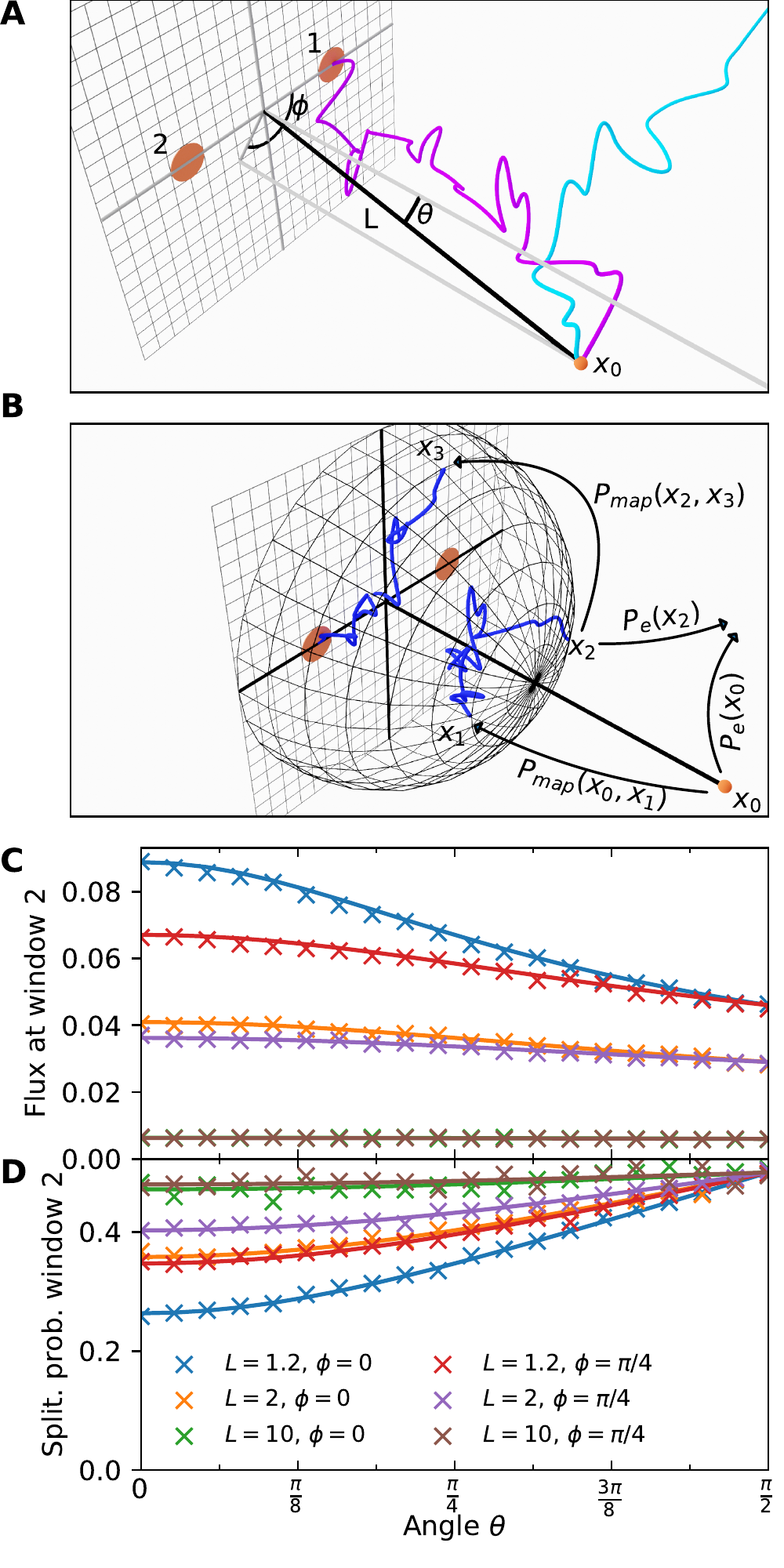}
    \caption{Fluxes to two windows located on a plane. (A) Simulation scheme: Brownian particles are released from the source $\x_0$ at a distance $L$ from the origin located on the plane. A trajectory is either absorbed by window 1 or 2 (magenta trajectory), or escapes to infinity (trajectory in cyan). (B) The position of a particle released by the source at $\x_0$ is mapped to the boundary of an imaginary half-sphere of radius $R$ enclosing the windows (black mesh), via the mapping probability distribution $P_{map}(x, y)$ given in Eq.~\ref{eq:MappingHalfspace}. Particles perform Brownian motion inside the half-sphere until they are absorbed by a window or they leave the half-sphere with radius $R'>R$ (green mesh) upon which they are mapped back again (see algorithm listing below). (C) Flux through window 2 vs the source zenith angle $\theta$, the azimuthal angle $\phi$ and the distance $L$: analytical solution \ref{eq:halfspacefluxn=2} (solid lines) compared to simulation results (cross markers). (D) Splitting probability for a particle to hit window 2 conditional on hitting either one of the two windows.}
    \label{fig:figure1}
\end{figure}
To determine the range of validity of the asymptotic formula, we designed a hybrid-stochastic simulation algorithm. This algorithm avoids the explicit simulation long trajectories with large excursions and thus it circumvents the need for an arbitrary cutoff distance for our infinite domain. The algorithm consists of mapping the source position $\x_0$ to a half-sphere containing the absorbing windows (Fig. \ref{fig:figure1}A). This mapping is defined in Appendix A. Inside the sphere, we run Brownian simulations, until the particle is absorbed or exits through the sphere surface. The detailed algorithm consists of the following steps, as illustrated in Fig. \ref{fig:figure1}B:
\begin{enumerate}
    \item The source releases a particle at position $\x^{t=0}=\x_0$.
    \item If $|\x^t|>R'$, we map the particle's position to the surface of the sphere $S(R)$, using $P_{map}$~(Eq. \ref{eq:MappingHalfspace} in appendix~\ref{a:mapping}). Note that there is a finite probability for the particle to escape to infinity upon which we terminate the trajectory.
    \item We use the Euler-Maruyama scheme to perform a Brownian step by calculating
    \beq
    \x^t=\x^{t-\Delta t}+\sqrt{2D\Delta t}\boldmath{R}^t,
    \eeq
   where $\boldmath{R}^t$ is a vector of standard normal random variables.
    \item We check whether the particle crosses any reflective boundary. If so, we repeat step 3 after discarding the new position.
    \item When $|\x^{t}-\x_i|<\eps$ for any $i$ ($\x_i$ is the position of window $i$), we consider that the particle is being absorbed by window $i$ and terminate the trajectory. Otherwise we return to step 2.
\end{enumerate}
Note that the radius $R'>R$ is necessary to prevent frequent re-crossings of the sphere $S(R)$ and thereby enhances computational efficiency.\\
\subsection{Computing the fluxes for two windows}
To validate our simulation scheme and our analytical formula, we numerically evaluated equation~\ref{twowindows} and compared these results with the results of our stochastic simulations (Fig. \ref{fig:figure1}C-D) for the flux through window 2, $\Phi_2$, and the splitting probability $p_2=\frac{\Phi_2}{\Phi_1+\Phi_2}$ for a continuous zenith angle $\theta$, various source distance values $L=d(0,\x_0)$ and the azimuthal angle $\phi$ either zero or $\pi/4$. As $L$ increases, the slitting probability increases quickly to .5, suggesting that for source distances greater than 10 times the distance between the two windows determining the direction of the source becomes impossible.
\subsection{ Computing the fluxes for three windows}
\begin{figure}[http!]
    \centering
    \includegraphics[scale=0.7]{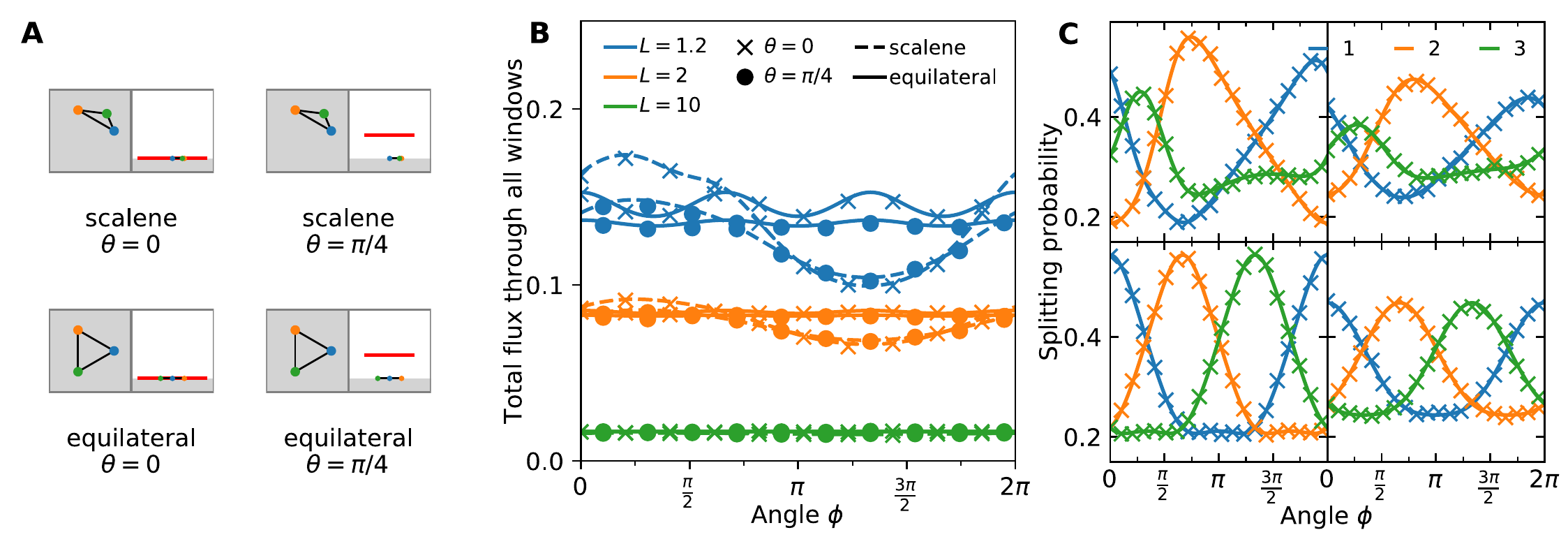}
    \caption{Fluxes to three windows located on the boundary of half-space. (A) Window configurations in the plane. The windows are arranged either in an equilateral ($\alpha=2\pi/3$, $\beta=-2\pi/3$) or a scalene ($\alpha=2\pi/3$, $\beta=0.4$) triangle, with a circumcircle radius of one. The source is kept at a distance $L$ and we varied its azimuthal angle $\phi$ continuously between $0$ and $2\pi$ while the zenith angle $\theta$ equals $0$ or $\pi/4$. (B) Total summed flux through all windows for the four different configurations as a function of the azimuthal angle $\phi$ and the source distance $L$. (C) Splitting probability for particles to hit a given window.}
    \label{fig:figure2}
\end{figure}

For three windows, we first computed the total flux through all windows $\Phi_t=\Phi_1+\Phi_2+\Phi_3$, depending on the window configuration and the position of the source.
Here, the source is located at a distance $L$ from the origin and we vary its position on a
circle in a plane parallel to the boundary. Therefore, the distance perpendicular
to this circle is $L\sin\theta$ and its radius is $L\cos\theta$, where
$\theta$ is the angle subtended by the plane and the source
position vector $\mathbf{OS}$ (Fig. \ref{fig:figure2}A). We investigated two types of window configurations: a scalene (non-symmetric) and an equilateral triangle (with an edge length of $\sqrt{3}/2$). The total flux through all three windows over the in-plane
source position angle $\phi$ for $\theta=0$ and $\pi/4$, reveals that for a source positioned very close to the windows, $L=1.2$, about $15\%$ of the flux is captured by the windows with the remainder escaping to infinity (Fig. \ref{fig:figure2}B). For a source far away from the windows ($L=10$), the captured flux decreases to about $2\%$. Neither the window configuration nor the angle $\theta$ has much influence on the total flux except when the source is very close. We again found excellent agreement between our analytical and simulation results for both the total flux and for the splitting probabilities $p_i=\frac{\Phi_i}{\Phi_1+\Phi_2+\Phi_3}$, $i=1...3$ (Fig. \ref{fig:figure2}C). The peaks in the splitting probability indicate the in-plane angles $\phi$ at which the source is closest to the corresponding window.
\section{Triangulating the source position from the fluxes}\label{triangulating_source}
Finding the source when the measured fluxes through the windows $P_1$, $P_2$ and $P_3$ are given can be categorized into the general class of inverse problems. Contrary to the two-dimensional case~\cite{dobramysl2018reconstructing,dobramysl2018mixed}, with three windows, the sum of the fluxes $\Phi_1+\Phi_2+\Phi_3$ is strictly less than one because particles can escape to infinity in three dimensions. For this reason,the flux values provide separate and independent pieces of information which allows us to recover the source position with at least three windows. In two dimensions, due to the recurrence property of Brownian motion the sum over all fluxes has to necessarily be one. Therefore, they are linearly dependent and at least three windows are required for the source reconstruction, albeit only two coordinates need to be recovered.

We recall that equations \ref{fluxesN=3} show that when the three fluxes are given, the source is located at the intersection of three overlapping spherical surfaces (to first order), the intersection of which yields the position $\x_0$. The position $\x_0$ only appears as the argument of the Neumann-Green's function $\mathcal{N}(\x,\y)$. In the absence of an analytical inverse of $\mathcal{N}(\x,\y)$ we proceed numerically. Therefore, when the distance between any window and the source, and the distances between the windows are large compared to $\eps$, we can use the leading order approximation to recover $\x_0$. In this case an analytical solution exists and is computed as follows.

Without loss of generality, we assume that the window positions are $\x_1=(0, 0, 0)$, $\x_2=(d, 0, 0)$ and $\x_3=(e, f, 0)$ (i.e. the windows all lie in the x-y plane, window 1 is at the origin and window 2 is on the x-axis). Then, using the leading order from the expansion of the fluxes in eq~\ref{fluxesN=3}, we have three non-linear equations for the location of the source
\beq
\gamma_1^2 &= (x_0^{(1)})^2 + (x_0^{(2)})^2 + (x_0^{(3)})^2\\
\gamma_2^2 &= (d-x_0^{(1)})^2 + (x_0^{(2)})^2 + (x_0^{(3)})^2\\
\gamma_3^2 &= (e-x_0^{(1)})^2 + (f-x_0^{(2)})^2 + (x_0^{(3)})^2,
\eeq
where $\gamma_i=\frac{2\eps}{\pi\Phi_i}$. Solving for the coordinates of $\x_0$ and requiring that $x_0^{3}>0$, we arrive to the analytical solution
\beq
x_0^{(1)}&=&\ds \frac{d^2+\gamma_1-\gamma_2}{2d}\\
x_0^{(2)}&=&\ds \frac{1}{2df}\left[d(e^2+f^2+\gamma_1-\gamma_3)-e(d^2+\gamma_1-\gamma_2)\right]\\
x_0^{(3)}&=& \ds \frac{1}{2df}\left[(e^2+f^2)(\{\gamma_1-\gamma_2\}^2-d^4)+2de(e^2+f^2+\gamma_1-\gamma_3)(d^2+\gamma_1-\gamma_2)\right.\\
&& \left.-d^2(e^4+f^4+[\gamma_1-\gamma_3]^2+2e^2[f^2+2\gamma_1-\gamma_2-\gamma_3]-2f^2[\gamma_2+\gamma_3])\right]^{1/2}.
\eeq

\begin{figure}
  \centering
  \includegraphics[scale=1]{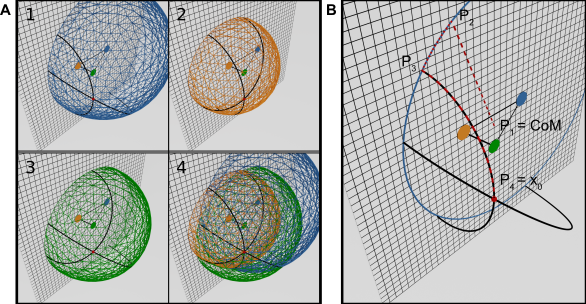}
  \caption{(A) Reconstruction of the source from the intersection of the surfaces defined by Eqs.~\ref{exactN=3} (color corresponding to the originating window's color). The source $\x0= (1, -1, 0)$(red dot) is recovered from the intersection of all three surfaces for the flux values $\phi_1\approx 0.03068, \phi_2\approx 0.04947$ and $\phi_3\approx 0.0358$. (B) Curve following algorithm schematic for windows located on the $x-y$ plane. The red dashed lines indicate the path the algorithm traces, starting close to the origin and ending at the source position. The blue circle is the intersection of the surface $S_k$ with the $x-y$ plane. The individual segments are labelled with the corresponding steps in the algorithm.}
  \label{curveTracing}
\end{figure}

We next develop a numerical procedure to find the position of the source $\x_0$ which is valid to any order in $\eps$. We introduce error function
\beqq
F_i(\x_0)=\theta_\eps\Phi_i+\sum_{j\neq i}\mathcal{N}(\x_i,\x_j)\Phi_j-2\pi\mathcal{N}(\x_i,\x_0)=0.
\eeqq
To find the position of the source $\x_0$ from the measured fluxes $\Phi_i$, $i=1...N$, we need to invert Eqs.~\ref{sysMatrix} (or Eqs.~\ref{sysMatrix2} in the case of a ball), together with Eq.~\ref{totalflux}. Each of the equations in Eqs.~\ref{sysMatrix} describes a non-planar surface $S_{i}$ in three dimension, corresponding to window $i$ and intersecting the half-plane (in the case of the windows located on the half-plane) or the unit ball (in the case of the windows located on the ball). Each pair of surfaces $S_i$ and $S_j$ intersect, forming three-dimensional curves $C_{ij}$ and all of these curves intersect at the location of the source $\x_0$. Hence we need at least three windows to find the source position. In the case of $N>3$ windows, we shall simply choose a combination $k$, $l$ and $m$ of three fluxes from the $N$ available.\\
The most straightforward way to find $\x_0$ would be to numerically find the global minimum of $E_{klm}(\x)=|F_k(\x)|+|F_l(\x)|+|F_m(\x)|$. However, this leads to issues due to many shallow local minima formed by the curves $C_{ij}$ that trap minimization algorithms. As an alternative, we find and follow one curve $C_{ij}$ to the root of all three conditions $F_k=0$, $F_l=0$ and $F_m=0$ with the following algorithm. We proceed with windows located on the $x-y$ plane (Fig. \ref{curveTracing}).
\subsection{Triangulating the source position when the window are on a plane} \label{ss:trianplan}
\begin{enumerate}
\item Define the initial step size $\Delta x$, the starting point $\mathbf{P}_1=(\Delta x, \Delta x, 0)$ and the error tolerance $\nu$.
\item Calculate the gradient vector $\v_1=\frac{dF_k}{d\x}(\mathbf{P}_1)$ and its projection on the $x-y$ plane $\tilde{\v}_1=\v_1-(\v_1\cdot\mathbf{e}_z)\mathbf{e}_z$. Find the root $\mathbf{P}_2=\mathbf{P}_1+t\tilde{\v}_1$ where $t$ is such that $F_k(\mathbf{P}_1+t\tilde{\v}_1)=0$, using Newton's algorithm.
\item Calculate the gradient vector $\v_2=\frac{dF_l}{d\x}(\mathbf{P}_2)$ and its projection to the $x-y$ plane $\tilde{\v}_2=\v-(\v_2\cdot\mathbf{e}_z)\mathbf{e}_z$. Find the root $\mathbf{P}_3=\mathbf{P}_2+t\tilde{\v}_2$ where $t$ is such that $F_l(\mathbf{P}_2+t\tilde{\v}_2)=0$ using Newton's algorithm.
\item Calculate the error on $F_k$ when we moved to $F_l$ by $e_{kl}=|F_k(\mathbf{P}_3)|$. If $e_{kl} > \nu$, go to step 2. Otherwise, we have now found the intersection $\mathbf{P}_3$ between the curve $C_{kl}$ and the $x-y$ plane within tolerance $\nu$ and can move on to tracing the curve $C_{kl}$.
\item Set $\y_0=\mathbf{P}_3$ and $\y_1=\mathbf{P}_3+dx\mathbf{e}_z$.
\item Calculate the gradient vector $\v_1=\frac{dF_k}{d\x}(y_1)$. Find the root $y_2=\y_1+t\v_1$ where $t$ is such that $F_k(\y_1+t\v_1)=0$, using Newton's algorithm.
\item Calculate the gradient vector $\v_2=\frac{dF_l}{d\x}(y_2)$. Find the root $y_3=\y_2+t\v_2$ where $t$ is such that $F_l(\y_2+t\v_2)=0$, using Newton's algorithm.
\item Calculate the error on $F_k$ when we moved to $F_l$ by $E_{kl}=|F_k(\y_3)|$. If $E_{kl} > \nu$, go to step 6.
\item Set $\w=\y_3-\y_0$, $\y_0=\y_3$ and $\y_1=\y_0+\w$. Calculate the error on $F_m$ via $E_{m}=|F_m(\y_0)|$. If $E_{kl} > \nu$, go to step 6. Otherwise, we have found the source location at $\mathbf{P}_4=y_0$ within tolerance $\nu$.
\end{enumerate}
This algorithm starts close to the origin and proceeds to find the intersection of the $C_{kl}$ curve with the $x-y$ plane. It then traces the curve $C_{kl}$ until it finds its intersection with the $S_m$ surface, where the source is located. We implemented this algorithm in python and the result is shown in Fig.~\ref{fig:figure2}D, where we also show the three surface associated to
\subsection{Triangulating the source position when the window are on a ball}
The case of windows on the ball is similar to the case of the windows on a plane:
\begin{enumerate}
\item Define the initial step size $\Delta x$. Calculate the center of mass of the windows $x_m=\sum_i\x_i/N$ and its projection onto the unit ball $\tilde{\x}_m=\x_m/|\x_m|$. Define the starting point $\y_0=[\x_m+(\Delta x, \Delta x, 0)]/|\x_m+(\Delta x, \Delta x, 0)|$ and the error tolerance $\nu$.
\item Calculate the gradient vector $\v_1=\frac{dF_k}{d\x}(y_0)$. Define the geodesic $\mathbf{G}(t)=[\y_0+t\v_1]/|\y_0+t\v_1|$ and find the root $\y_1=\mathbf{G}(\tilde{t})$ where $\tilde{t}$ is such that $F_k(\tilde{t})=0$, using Newton's algorithm.
\item Calculate the gradient vector $\v_2=\frac{dF_l}{d\x}(y_1)$. Define the geodesic $\mathbf{G}(t)=[\y_1+t\v_2]/|\y_1+t\v_2|$ and find the root $\y_2=\mathbf{G}(\tilde{t})$ where $\tilde{t}$ is such that $F_l(\tilde{t})=0$ using Newton's algorithm.
\item Calculate the error on $F_k$ when we moved to $F_l$ by $e_{kl}=|F_k(\y_2)|$. If $e_{kl} > \nu$, go to step 2. Otherwise, we have now found the intersection between the curve $C_{kl}$ and the unit ball within tolerance $\nu$ and can move on to tracing the curve $C_{kl}$.
\item Set $\y_0=\y_2$ and $\y_1=(1+dx)\y_0$.
\item Calculate the gradient vector $\v_1=\frac{dF_k}{d\x}(y_1)$. Find the root $y_2=\y_1+t\v_1$ where $t$ is such that $F_k(\y_1+t\v_1)=0$ using Newton's algorithm.
\item Calculate the gradient vector $\v_2=\frac{dF_l}{d\x}(y_2)$. Find the root $y_3=\y_2+t\v_2$ where $t$ is such that $F_l(\y_2+t\v_2)=0$ using Newton's algorithm.
\item Calculate the error on $F_k$ when we moved to $F_l$ by $E_{kl}=|F_k(\y_3)|$. If $E_{kl} > \nu$, go to step 6.
\item Set $\w=\y_3-\y_0$, $\y_0=\y_3$ and $\y_1=\y_0+\w$. Calculate the error on $F_m$ via $E_{m}=|F_m(\y_0)|$. If $E_{kl} > \nu$, go to step 6. Otherwise, we have found the source location at $y_0$ within tolerance $\nu$.
\end{enumerate}
\subsection{Sensitivity analysis}
\begin{figure}
    \centering
    \includegraphics[scale=0.8]{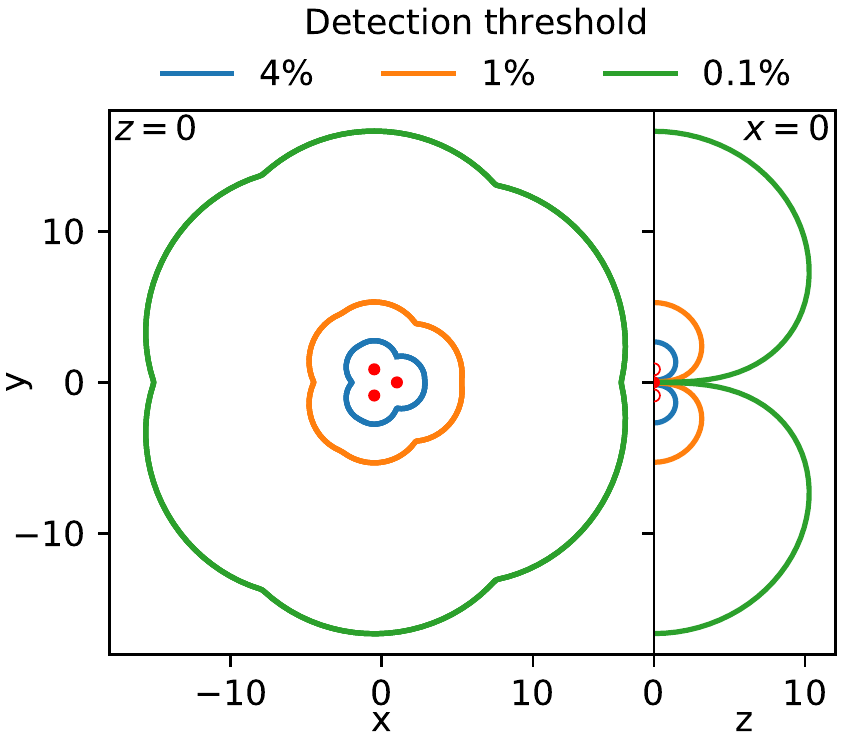}
    \includegraphics[scale=0.6]{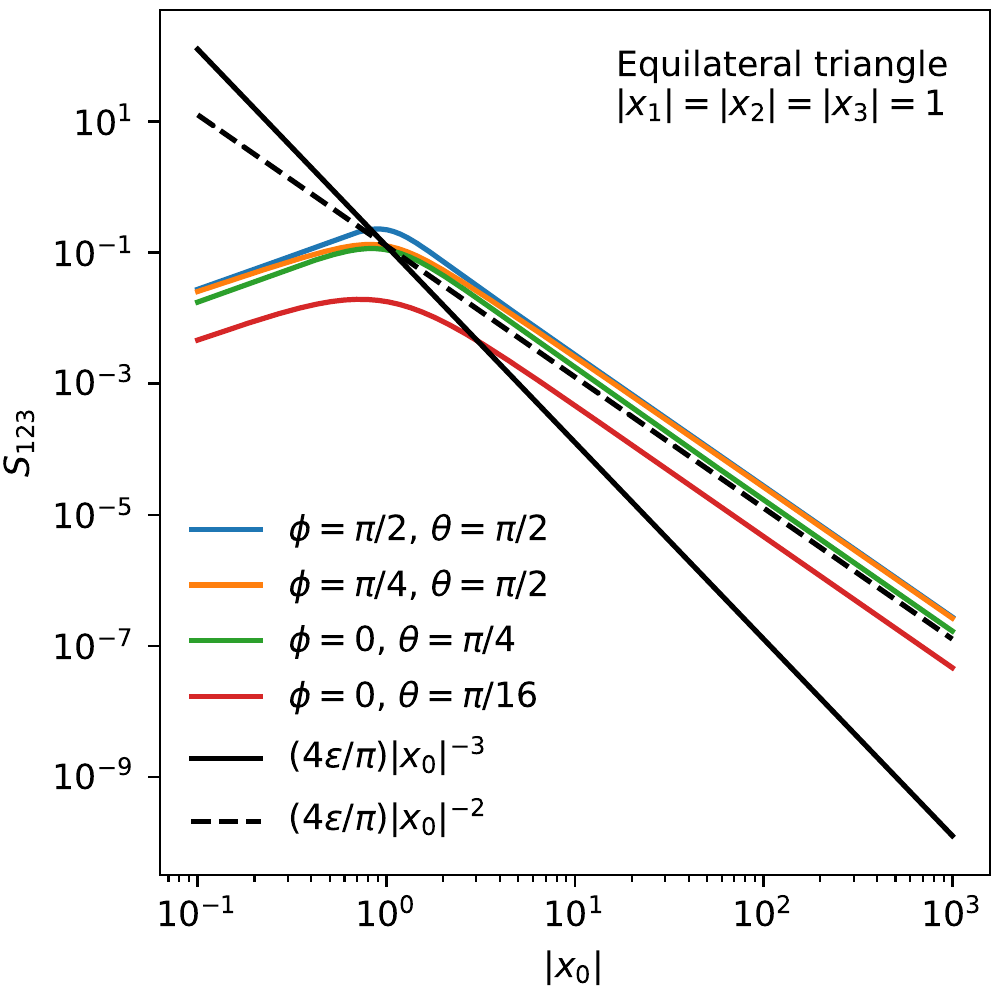}
    \caption{{\bf Sensitivity of detecting the source position from
      Eq.~(\ref{eq:S123})}. (Left) for a ball with three windows arranged as an
      equilateral triangle on a geodesic. The detection contours is in
      the plane that contains all 3 windows (left) and in plane perpendicular to the window plane (right), for three different detection thresholds (1\%, 0.1\% and 0.01\%). (Right) The sensitivity decays with $1/|x_0|^2$ for the source position $x_0$.}
    \label{fig:figure3}
\end{figure}

To explore how far away the source can be recovered, we introduce a sensitivity function, which is expressed as the differences between all splitting probability computed from the fluxes
\beq\label{eq:S123}
S_{123}(\x_0;\x_1,\x_2,\x_3)=\max\{&|P_1(\x_0)-P_2(\x_0)|, \nonumber
  &|P_2(\x_0)-P_3(\x_0)|,\\&|P_3(\x_0)-P_1(\x_0)|\},
\eeq
where $\x_0$ is the position of the source and $\x_i$, $i=1, 2, 3$ are the positions of the three windows on $\p B_a$. The cost function $S_{123}$  describes the maximum absolute imbalance between the fluxes through the windows. Fig.~\ref{fig:figure3}A shows the contours of this function for three windows arranged in an equatorial equilateral triangle in a slice through the $z=0$ and $x=0$ planes at three
different threshold levels. Notably, the distance at which directions
can still be discerned is approximately an order of magnitude less for
any given threshold compared to the equivalent situation in two dimensions ~\cite{dobramysl2018reconstructing}. Indeed, using the dipole expansion for a source located far away $|\x_0|\gg 1$, $f(\x_0;\x_1,\x_2,\x_3)\approx \frac{C}{|\x_0|^2}$, where $C>0$ is constant and $\hat{\x_0}=\frac{\x_0}{|\x_0|}$. Fig.~\ref{fig:figure3}B illustrates this decay.
\subsection{Region of uncertainty to recover the source}\label{region_of_uncertainty}
To account for the possible error in the reconstructed source location due to measurement fluctuations in the window fluxes $\Phi_i$, we define an uncertainty region $R_{unc}$ and we will estimate its volume $V_{unc}$. This region contains the location of the source position and its size represents the positional uncertainty stemming from the fluctuations in the fluxes. A small region $R_{unc}$ indicates a very accurate reconstruction, while a large $R_{unc}$ means high uncertainty in at least one direction. Here, we present a method to construct this region, as intersections of parallelepipeds. We describe the perturbation due to measurement error as
\beq
\tilde{\Phi}_i=\Phi_i+\eta,
\eeq
where $\eta\ll\Phi$ is an additive constant representing the error.

\begin{figure}
    \centering
    \includegraphics[scale=0.7]{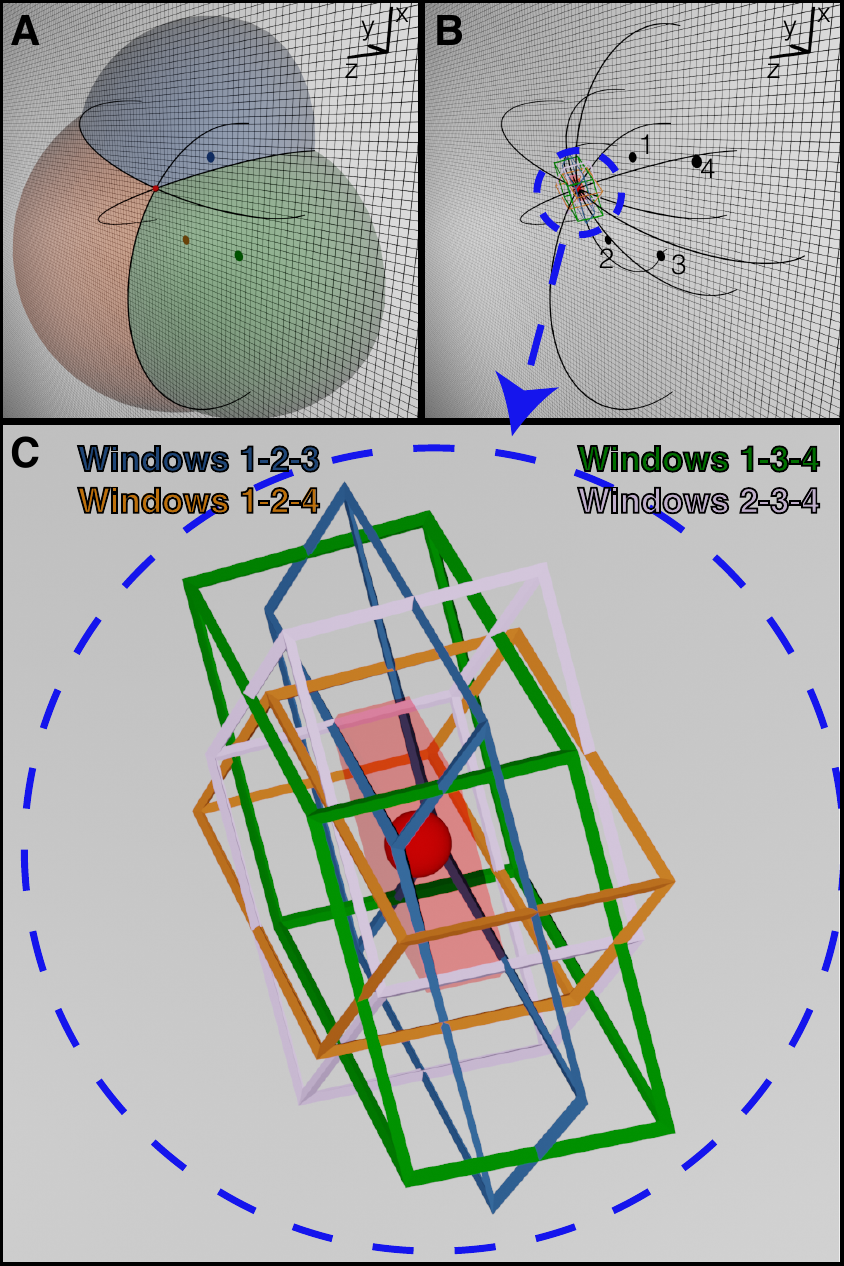}
    \caption{{\bf Triangulation of the source position with three out of four windows}. (A) Triangulation using fluxes from three windows only. (B) A further window yields additional redundant intersection lines. (C) Enlargement of the area around the source in (B). Each combination of three windows defines a volume (parallelepiped) around the source position, computed from combining 3 out of 4 windows (various colors). The intersection of these volumes defines the uncertainty volume $V_{unc}$ (shaded red).}
    \label{fig:figure3bis}
\end{figure}
Intuitively, to first order in $\eps$, the fluxes decay as a power law of the distance to a particular window. Hence, we start with a procedure valid to leading order in $\eps$ and similar to our simplified source reconstruction in section~\ref{triangulating_source}. The source is located is on a sphere centered around the window $i$ and with a radius $\tilde{R}_i=2\eps/(\pi\Phi_1)$. Therefore, the location of the source varies according to $-d\tilde{R}_i/d\Phi_i=2\eps/(\pi\Phi_i^2)$ along the radial vector $\x_0-\x_i$. The complete expression for the error vector associated with window $i$ is then given by
\beq
\vec{e}_i=\eta_i\frac{2\eps}{\pi\Phi_i^2}\frac{\x_0-\x_i}{|\x_0-\x_i|}.
\eeq
For three windows, the vectors $e_1$, $e_2$ and $e_3$ describes a
parallelepiped, the volume of which represents the measure of location
uncertainty. As we shall describe below, the volume is inhomogeneous, it depends both on the location of the source and the particular arrangement of the windows.

The precise, numerical procedure is as follows: The $i-$th coordinates of the reconstructed source position is given for the window indices $k$, $l$ and $m$ (three windows out of the $N$ available), by a Taylor expansion using the flux coordinate system, mentioned in subsection \ref{ss:explicit}:
\beq
\begin{split}
  \tilde{x}_{0}^i(\tilde{\Phi}_k, \tilde{\Phi}_l, \tilde{\Phi}_m) &= \tilde{x}_{0}^i(\Phi_k+\eta, \Phi_l+\eta, \Phi_m+\eta) \\
  &=\tilde{x}_{0}^i(\Phi_k, \Phi_l, \Phi_m)+\eta\left(\frac{\p x_{0}^i}{\p\Phi_k}+\frac{\p x_{0}^i}{\p \Phi_l}+\frac{\p x_{0}^i}{ \p \Phi_m}\right) + O(\eta^2)\\
  &=x^i_{0}+\eta(E^{(k,l,m)}_{1i}+E^{(k,l,m)}_{2i}+E^{(k,l,m)}_{3i}) + O(\eta^2),
\end{split}
\eeq
where we used that $\tilde{\x}_{0}(\Phi_k, \Phi_l, \Phi_m)=\x_0$ and the error matrix is defined as $E^{(k,l,m)}_{ij}= \frac{\p x_0^j}{d\Phi_i}$. Therefore, to evaluate the uncertainty region, we compute the Jacobian
\beq
J_{ij}=\frac{\p x_0^j}{\p \Phi_i},
\eeq
for three  fluxes $k$, $l$ and $m$. The linear uncertainty vectors $E^{(k,l,m)}_1$, $E^{(k,l,m)}_2$ and $E^{(k,l,m)}_3$ span a parallelepiped $P_u$ at the location of the source $\x_0$. Therefore, the volume of uncertainty for the source reconstruction is the volume of this parallelepiped
\beq
V_{\mathrm{unc}}^{(k,l,m)}(\eta)=\eta/|\det(J_{ij})|.
\eeq
Because the choice of the three windows $k$, $l$ and $m$ is arbitrary, we define the total volume of uncertainty $V_{unc}(\eta)$ as the volume of the geometric intersection of all parallelepipeds generated by the possible combinations of any three window fluxes from the $N$ available. The intersection of parallelepipeds is illustrated in Fig. \ref{fig:figure3bis}:
Using three windows only (Fig. \ref{fig:figure3bis}A) we reconstruct the source location $\x_0$ using the algorithm introduced in section \ref{ss:trianplan}. When adding a fourth window, we have four possible combinations of three from which the source can be reconstructed. Thus we obtain six curves from the intersection of the four surfaces (Fig. \ref{fig:figure3bis}B). The resulting four parallelepipeds are displayed in Fig. \ref{fig:figure3bis}C together with their geometric intersection (red volume).

\begin{figure}[http!]
    \centering
    \includegraphics{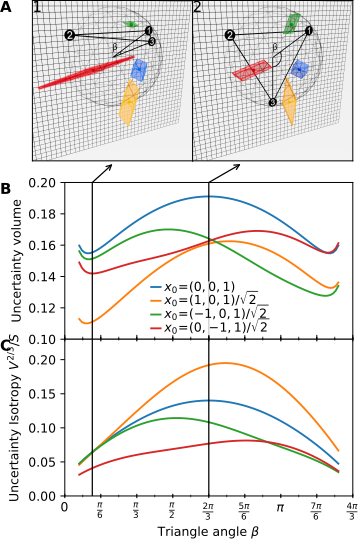}
    \caption{Uncertainty of the source recovery depends on the its location. (A) Volumes of uncertainty (colored regions) for four different position of the source. We show the volumes for two different configurations of three windows: (1) a scalene triangle and (2) an equilateral triangle. (B) Measured volume of uncertainty for different source positions vs the triangle angle $\beta$ ($\beta=0$ corresponds to window3 overlapping with window 1 while $\beta=2\pi/3$ corresponds to window 3 overlapping with window 2). (C) Uncertainty isotropy (isoperimetric ratio) as a function of the source position and the triangle angle $\beta$.}
    \label{fig:uncertainty_inhomog}
\end{figure}
\begin{figure}[http!]
    \centering
    \includegraphics{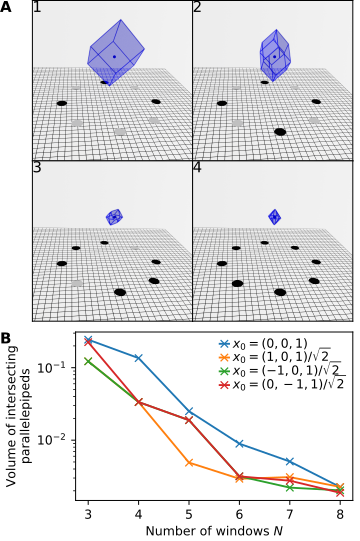}
    \caption{Uncertainty is reduced by the number of windows. (A) Three-dimensional display of the total volume of uncertainty (defined as the intersection of all parallelepipeds from all combinations of three windows for (1) three, (2) four, (3) six and (4) eight windows. (B) The total uncertainty volume as a function of the number of windows for four different source positions.}
    \label{fig:figure4}
\end{figure}

The region of uncertainty $R_{unc}$ and its volume $V_{unc}(\eta)$ strongly depend on the position of the source relative to the windows. This is illustrated in Fig.\ref{fig:uncertainty_inhomog}A where we show the $R_{unc}$ for four different source positions and two different window configurations (a scalene and an equilateral triangle). To further quantify the uncertainty volume, we vary the triangle angle $\beta$ and compute the volume and the isoperimetric ratio $S/V^{2/3}$, where $S$ is the surface and $V$ the volume (Fig.~\ref{fig:uncertainty_inhomog}B,C). The parallelepipeds can be highly elongated (Fig.~\ref{fig:uncertainty_inhomog}C). Interestingly, the minimum of the isoperimetric ratio (i.e. the triangle angle $\beta$ at which $R_{unc}$ is most isotropic) strongly depends on the source position (Fig.\ref{fig:uncertainty_inhomog}C).\\
When the number of windows $N$ is larger than three, there are  $N!/(3![N-3]!)$ combinations of the $N$ error vectors $e_i$ (see Fig.~\ref{fig:figure4}A for illustrations of $N=3, 4, 6$ and $8$). The volume of uncertainty decreases super-exponentially when the number of windows $N$ increases (see Fig.~\ref{fig:figure4}B).
\section{Concluding remarks}\label{discussion}
In this manuscript we presented a general method to compute the steady-state fluxes of Brownian particles to narrow windows located on a surface. We developed a hybrid stochastic simulation approach, which consists of replacing random walks between the point source and a window by mapping the source position to an imaginary surface (a half-sphere in the case of half-space and an entire sphere in the case of a ball, both in three dimensions), followed by a stochastic step where the Brownian trajectories are simulated in a small neighborhood of the surface. The analytical part of the method is based on computing the asymptotic solution of Laplace's equation using the Neumann-Green's function and matched asymptotics. The analytical relation between the flux expressions and the location of the source that we found leads to a reconstruction procedure of the source from measured fluxes.

In addition, this approach allows us to estimate how measurement fluctuations in the fluxes can be compensated by increased number of narrow windows. The uncertainty is represented as the volume of the Jacobian matrix for any three windows. By considering the combinatorics of any three windows out of $N$ (binomial $C^3_N$), the uncertainty corresponds to the intersection of a large number of parallelepipeds (see subsection~\ref{region_of_uncertainty}). Finding the exact decay of the uncertainty volume with the number of windows remain an open question.\\
Note that the present approach can be extended to the case where the diffusion particles can be destroyed with a uniform killing rate $k(\x)=k$, that represent how cues can be degraded or get lost between the source of the windows \cite{HMS2005}, leading to an exponential decaying distribution.\\
This work was motivated by our wish to understand how cells can accurately identify the position of a gradient source in three dimensions.  For example, it remains unclear how neurons in the brain orient and navigate toward their final destination~\cite{chedotal,kolodkin2011mechanisms}. Even if the main molecular players have been identified, the physical mechanism that converts the external flux into a series of commands that generate the neuronal path is unclear. Especially the first step, which consists of reading an external gradient field, and internalizing this information at the growth cone level to determine when to grow or to stop at a given position remains, to be understood. The present study demonstrates that at least three receptors are sufficient to triangulate the position of the source and any additional one adds redundancy to increase the precision of the source localisation. Future works should consider the case of multiple sources in integrating the external signal.

\section*{Acknowledgements}
U.D. was supported by a Herchel Smith Postdoctoral Fellowship and acknowledges core funding by the Wellcome Trust (092096) and CRUK (C6946/A14492).

\section{Appendixes}\label{appendix0}
\subsection{Explicit Green's function mapping for a half-sphere on a reflecting plane}\label{a:mapping}
The mapping of a particle at a position $|\x|>R$ to the surface of the half-sphere with radius $R$ is given by the diffusive flux through this surface with absorbing boundary conditions. Therefore, we need to construct the Green's function for the infinite domain $\mathbb{R}^3_+-B(R)$ with Dirichlet boundary conditions at $\p B(R)$:
\beq
    \begin{split}
        -\Delta G(\x, \y) &= \delta(\x-\y) \quad \text{for} \quad \x\in\mathbb{R}^3_+\backslash B(R) \\
        \frac{\p G}{\p \n}(\x,\y) &= 0 \quad \text{for} \quad \x\in\p\Omega\\
        G(\x,\y) &= 0 \quad \text{for} \quad \x\in\p B(R)
    \end{split}
\eeq
Using the symmetries of the reflective half-plane and the sphere, we apply the method of images, starting with the Green's function for the absorbing ball in free space~\ref{eq:GreenFctMap3DBall}. The solution of this problem is
\beq
    \label{eq:GreenFctMap3DHalfspace}
    G(\x, \y) = -\frac{1}{4\pi}\left[\frac{1}{|\x-\y|} - \frac{|\x|}{R}\frac{1}{|\x-\y|\x|^2/R^2} + \frac{1}{|\x-\tilde{\y}|} - \frac{|\x|}{R}\frac{1}{|\x-\tilde{\y}|\x|^2/R^2}\right],
\eeq
where $\tilde{\y}$ is the reflected image of $\y$ through the plane. The mapping probability is thus
\beq \label{eq:MappingHalfspace}
    P(\x,\y)=\frac{1}{\sqrt{R^2+\rho^2-R\rho\kappa}^3} +
             \frac{1}{\sqrt{R^2+\rho^2-R\rho\tilde{\kappa}}^3}\,,
\eeq
where
\beq
\kappa=\cos(\theta-\theta')(\cos[\phi-\phi']+1)+\cos(\theta+\theta')(\cos[\phi-\phi']-1) \\ \tilde{\kappa}=\cos(\theta-\theta')(\cos[\phi-\phi']-1)+\cos(\theta+\theta')(\cos[\phi-\phi']+1),
\eeq
$\phi$ and $\phi'$ are the polar angles of $\x$ and $\y$ in the $x-y$ plane and $\theta$ and $theta'$ are their respective angles with the $z$-axis.

\subsection{Mapping the source for a ball in 3D} \label{a:mappingball}
The mapping of a particle released at a position $|\x|>R$ is given by the diffusive flux through an absorbing ball with radius $R$. Hence, we need to construct the Green's function for the infinite domain $\mathbb{R}^3/B(R)$ with Dirichlet boundary conditions at $\p B(R)$:
\beq
    \begin{split}
        -\Delta G(\x, \y) &= \delta(\x-\y) \quad \text{for} \quad \x\in\mathbb{R}^3\backslash B(R) \\
        G(\x,\y) &= 0 \quad \text{for} \quad \x\in\p B(R)
    \end{split}
\eeq
This is easily solved via the method of images (which is applicable in the Dirichlet case), and we arrive at
\beq
    \label{eq:GreenFctMap3DBall}
    G(\x, \y) = -\frac{1}{4\pi}\left[\frac{1}{|\x-\y|} - \frac{|\x|}{R}\frac{1}{|\x-\y|\x|^2/R^2}\right].
\eeq
The flux through the boundary is then given by
\beq
    \label{eq:pmap_3dball}
   \frac{\p G}{\p r}(r=R, \y)=\frac{1}{4\pi}\frac{\beta^2-1}{(1+\beta^2-2\beta\cos\gamma)^{3/2}},
\eeq
where $r=|\x|$, $\beta=|\y|/R$ and $|\x||\y|\cos\gamma=\x\cdot\y$.
Integrating over the ball yields
\beq \label{intfl}
\int_{\p B(R)}P(\x,\y)dS_{\x}=\beta^{-1}=\frac{R}{|\y|},
\eeq
which is the first passage probability for hitting the ball before escaping to infinity.
The probability distribution of hitting is thus obtained by normalizing the integral of the flux \ref{intfl}:
\beq\label{probas}
 P(\x,\y)=\frac{|\y|}{R}\frac{1}{4\pi}\frac{\beta^2-1}{(1+\beta^2-2\beta\cos\gamma)^{3/2}}
\eeq
A random new location on the ball of radius $R$ can then be generated by using the probability \ref{probas}.
\subsection{Exact Neumann-Green's function for the ball}\label{appendix1}
The Neumann's function  $\mathcal{\tilde{N}}(\x,\x_0)$ is the solution of Laplace's equation
\beq
\Delta \mathcal{\tilde{N}}(\x,\x_0)&=&-\delta(\x-\x_0) \hbox{ for } \x \in \mathbb{R}^3 \nonumber \\
\frac{\p \mathcal{\tilde{N}}}{\p n} (\x,\x_0) &=& 0 \hbox{ for } \x \in S_a=\p B_a.
\eeq
where $S_a$ is the sphere of the three-dimensional ball $B_a$ and the source point $\x_0 \in \rR^3-B_a$. The analytical expression of the Neumann function \cite{Lagache2017}is
\beq \label{Neumann}
\mathcal{\tilde{N}}(\x,\x_0)&=& \ds\frac{1}{4\pi |\x-\x_0|}+\frac{a}{4\pi |\x_0||x-\ds \frac{a^2 \x_0}{|\x_0|^2}|}\nonumber \\&+&
\ds{\frac{1}{4\pi a }\log\left( \ds\frac{\ds\frac{|\x_0||\x|}{a^2}\left(1-\cos(\theta)\right)}{\ds 1-\frac{|\x_0||\x|}{a^2} \cos(\theta)+\left(1+\left(\frac{|\x_0||\x|}{a^2}\right)^2-2\frac{|\x_0||\x|}{a^2} \cos(\theta)\right)^{\frac{1}{2}}}\right)}. \label{N}
\eeq
Note that when $\x$ and $\x_0$ are on the sphere $S_a$, $|\x_0|=|\x|=a$, we obtain the expression:
\beqq
\mathcal{\tilde{N}}(\x,\x_0)&=&\frac{1}{2\pi |\x-\x_0|}+ \frac{1}{4\pi a }\log\left(\frac{|\x-\x_0|}{2a+|\x-\x_0|}\right). \label{Neumann2}
\eeqq
The far field expansion for $|\x|\gg 1 $ is given by
\beq \label{decay}
\mathcal{\tilde{N}}(\x,\x_0)&\approx&  \frac{1}{4\pi |\x|} +\frac{3 \x .\x_0}{8\pi |\x|^3} +O(\frac{1}{|\x|^3})\\
\mathcal{\nabla \tilde{N}}(\x,\x_0)&\approx& O(\frac{1}{|\x|^2})
\eeq


\begin{thebibliography}{99}
\bibitem{wolpert1996one} L. Wolpert, One hundred years of positional information, Trends Genet., 12 (1996) 359--364.

\bibitem{Malherbe} G.~Malherbe and D.~Holcman, Stochastic modeling of gene activation and application to cell regulation, J. Theor. Biol. 271 (2010), 51--63.

\bibitem{Kasatkin2} V.~Kasatkin, A.~Prochiantz and D.~Holcman, Morphogenetic gradients and the stability of boundaries between neighboring morphogenetic regions (2007), Bull. Math. Biol. 70, 156--78.

\bibitem{kasatkin2008morphogenetic} V. Kasatkin, A. Prochiantz and D. Holcman, Morphogenetic gradients and the stability of boundaries between neighboring morphogenetic regions, Bull. Math. Biol. 70 (2008), 156--178.

\bibitem{reviewREingruber} J. Reingruber and D. Holcman, Computational and mathematical methods for morphogenetic gradient analysis, boundary formation and axonal targeting, Sem. Cell Dev. Biol. 35 (2014) 189-202.

\bibitem{Heinrich3} D. Arcizet, S. Capito, M. Gorelashvili, C. Leonhard, M. Vollmer, S. Youssef, S. Rappl and D. Heinrich, Contact-controlled amoeboid motility induces dynamic cell trapping in 3D-microstructured surfaces. Soft Matter 8 (2012) 1473-1481.

\bibitem{Kaupp1} U. B. Kaupp, and T. Str\"{u}nker, Signaling in Sperm: More Different than Similar. Trends Cell Biol. 2016 (2016), S0962.


\bibitem{BergPurcell} H. C. Berg and E. M. Purcell, Physics of chemoreception, Biophys. J. 20 (1977) 193.

\bibitem{Endres2008} R. G. Endres and N. S. Wingreen, Accuracy of direct gradient sensing by single cells, Proc. Nat. Acad. Sci. U.S.A. 105 (2008) 15749.

\bibitem{SIREV-biol} D. Holcman, and Z. Schuss, The narrow escape problem, SIAM Rev. 56, 213--257 (2013).

\bibitem{coombs} D. Coombs, R. Straube and M. Ward. Diffusion on a sphere with localized traps: Mean first passage time, eigenvalue asymptotics, and Fekete points. SIAM Journal of Applied Mathematics, 70:302-332 DOI:10.1137/080733280

\bibitem{ward1} {M. J. Ward, W. D. Henshaw, J. B. Keller, Summing logarithmic expansions for singularly perturbed eigenvalue problems, SIAM J. Appl. Math. 53 (1993) 799-828.}

\bibitem{ward2} M. I. Delgado, M. J. Ward, D. Coombs, Conditional mean first passage times to small traps in a 3-D domain with a sticky boundary: applications to T cell searching behavior in lymph nodes, Multiscale Model. Simul. 13 (2015) 1224–1258.

\bibitem{ward3} V. Kurella, J. C. Tzou, D. Coombs, M. J. Ward, Asymptotic analysis of first passage time problems inspired by ecology, Bull. Math. Biol. 77 (2015) 83–125.


\bibitem{JPA2017} D.~Holcman and Z.~Schuss, 100 years after Smoluchowski: stochastic processes in cell biology. J. Phys. A: Math. Theor. 50 (2017), 093002.

\bibitem{Holcman2015} D. Holcman and Z. Schuss, Stochastic Narrow Escape in Molecular and Cellular Biology: Analysis and Applications, Springer (2015).

\bibitem{dobramysl2018mixed} U. Dobramysl and D. Holcman, Mixed analytical-stochastic simulation method for the recovery of a brownian gradient source from probability fluxes to small windows, J. Comp. Phys. 355 (2018), 22--36.

\bibitem{dobramysl2018reconstructing} U. Dobramysl and D. Holcman, Reconstructing the gradient source position from steady-state fluxes to small receptors, Sci. Rep. 8 (2018), 941.

\bibitem{Flegg2011} M. B. Flegg, S. J. Chapman and R. Erban, The two-regime method for optimizing stochastic reaction–diffusion simulations, J. Royal. Soc. Inter. 9 (2011), 859-868.

\bibitem{Franz2013} B. Franz, M. B. Flegg, S. J. Chapman and R. Erban, Multiscale reaction-diffusion algorithms: PDE-assisted Brownian dynamics, SIAM J. Appl. Math. 73 (2013), 1224-1247.

\bibitem{Smith2018} C. A. Smith, C. A. Yates, Spatially extended hybrid methods: a review, J. Royal Soc. Inter. 15 (2018), 20170931.

\bibitem{Holcman2008_1} D.~Holcman and Z.~Schuss, Diffusion through a cluster of small windows and flux regulation in microdomains, Phys. Lett. A 372 (2008), 3768--3772.

\bibitem{Schuss:Book} Z.~Schuss, Diffusion and Stochastic Processes. An Analytical Approach, Springer-Verlag, New York, NY, 2009.

\bibitem{Jackson} J. D. Jackson, Classical Electrodynamics, 3rd Ed. (1999), John Wiley \& Sons, New York.

\bibitem{Lagache2017} T. Lagache, and D. Holcman, Extended narrow escape with many windows for analyzing viral entry into the cell nucleus, J. Stat. Phys. 166 (2017), 244-266.


\bibitem{HMS2005} D. Holcman, A. Marchewka and Z. Schuss, Survival probability of diffusion with trapping in cellular neurobiology, Phys. Rev. E 72 (2005), 031910.

\bibitem{chedotal} A. Chedotal, and L. J. Richards, Wiring the Brain: The Biology of Neuronal Guidance, Cold Spring Harb. Perspect. Biol. 2010 (2010), a001917.

\bibitem{kolodkin2011mechanisms} A. L. Kolodkin and M. Tessier-Lavigne, Mechanisms and molecules of neuronal wiring: a primer, Cold Spring Harb. Perspect. Biol. 2011 (2011), a001727.
\end{thebibliography}
\end{document}